\documentclass[11pt]{article}
\usepackage{amsfonts}
\usepackage{epsfig,amssymb,euscript}
\usepackage{amsmath,amscd}

\numberwithin{equation}{section}
\newcommand{\be}{\begin{eqnarray}}
\newcommand{\ee}{\end{eqnarray}}
\newcommand{\bea}{\begin{eqnarray}}
\newcommand{\eea}{\end{eqnarray}}
\newcommand{\ba}{\begin{array}}
\newcommand{\ea}{\end{array}}

\def\appendix#1{\addtocounter{section}{1}\setcounter{equation}{0}
\renewcommand{\thesection}{\Alph{section}}
\section*{Appendix \thesection\protect\indent \parbox[t]{11.15cm}{#1}}
\addcontentsline{toc}{section}{Appendix \thesection\ \ \ #1}}

\begin{document}

\begin{titlepage}
\vfill
\begin{flushright}
\end{flushright}
\vfill
\begin{center}
   \baselineskip=16pt
  {\Large\bf Euclidean $N=2$ Supergravity }
   \vskip 2cm
       Jan B. Gutowski$^1$, 
      and W. A. Sabra$^2$\\
 \vskip .6cm
      \begin{small}
      $^1$\textit{Department of Mathematics, King's College London.\\
      Strand, London WC2R 2LS\\United Kingdom \\
        E-mail: jan.gutowski@kcl.ac.uk}
        \end{small}\\*[.6cm]
  \begin{small}
      $^2$\textit{Centre for Advanced Mathematical Sciences and
        Physics Department, \\
        American University of Beirut, Lebanon \\
        E-mail: ws00@aub.edu.lb}
        \end{small}      
   \end{center}
\vfill

\begin{center}
{\bf Abstract}
\end{center}
Euclidean special geometry has recently been investigated in the context of Euclidean supersymmetric theories with 
vector multiplets. In the rigid case, the scalar manifold is described by affine special para-K\"ahler geometry while the 
target geometries of Euclidean vector multiplets coupled to supergravity are given by projective 
special para-K\"ahler manifolds. In this letter, we derive the Killing spinor equations of
Euclidean $N=2$ supergravity theories coupled to vector multiplets. These equations provide the starting point for finding general supersymmetric instanton solutions.
\end{titlepage}

\section{Introduction}

Special geometry was first discovered in the study of the coupling of $N=2$
supergravity to vector multiplets \cite{witvan84}. In recent years, this geometry has
provided an important ingredient in the understanding of 
non-perturbative structure in field theory, supergravity, string
compactifications (see for example: \cite{onebigref}), as well as in the
study and analysis of black hole physics \cite{blackdev}.  More recently,
the Euclidean version of special geometry has been investigated in the
context of Euclidean supersymmetric theories \cite{mohaupt1, mohaupt2,
mohaupt3}. The Euclidean versions of the special
geometries can be obtained from their standard counterparts by replacing $i$
by the object $e$ with the properties $e^{2}=1$ and $\bar{e}=-e$. In the
context of finding instanton solutions, this replacement was first done in 
\cite{gb} in the study of D-instantons in type IIB supergravity.
Geometrically, this change of $i$ into $e,$ effectively is the replacement
of the complex structure by a para-complex structure. Details on
para-complex geometry, para-holomorphic bundles, para-K\"{a}hler manifolds
and affine special para-K\"{a}hler manifolds can be found in \cite{mohaupt1}. 
In the rigid case, the scalar manifold is described by
affine special para-K\"{a}hler geometry. Starting from the general
five-dimensional vector multiplet action, the dimensional reduction over a
time-like circle was considered in \cite{mohaupt1}. The Euclidean action,
together with the supersymmetry transformations when expressed in terms of
para-holomorphic coordinates, are of the same form as their Minkowskian
counterparts.

In \cite{mohaupt3} the results of the rigid case were generalised by
considering the dimensional reduction of the five dimensional supergravity
theory of \cite{GST}. The dimensional reduction with respect to a time-like
and space-like direction, gives respectively the Euclidean and Lorentzian
theories in four dimensions. The bosonic action for both types of reductions
was obtained in \cite{mohaupt3}. The target geometries of Euclidean vector
multiplets coupled to supergravity are given by projective special 
para-K\"{a}hler manifolds \cite{mohaupt3}. In this work, we
complete the analysis of \cite{mohaupt3} and determine the associated Killing spinor equations. 
These will be a step in the direction of the classification of instanton solutions with
non-trivial gauge and scalar fields. We organise this work as follows. We
review the bosonic reduction \cite{mohaupt3} in section 2. This will
fix our notation, as well as the relation between the five and four
dimensional bosonic fields needed to study the reduction of the Killing spinor equations
from five to four dimensions. Section 3 contains the
reduction of the Killing spinor equations. Section 4 describes how these equations can be rewritten
using an appropriate chiral decomposition, and recast into a $\epsilon$-complex form, or into an adapted
co-ordinate form. We conclude in section 5.

\section{Bosonic reduction and Special $\protect\epsilon $-K\"{a}hler
Geometry}

In this section we review the bosonic reduction of the five
dimensional supergravity theory \cite{mohaupt3}. The Lagrangian of the five
dimensional theory is given by \cite{GST}
\footnote{this is related to the original Lagrangian via the following identifications:
\par
\bigskip
\par
\begin{equation}
{\mathcal{F}}^{i}\rightarrow \frac{6^{1/6}}{2}{\mathcal{F}}^{i},\text{ \ \ \
\ \ \ }h^{i}\rightarrow 6^{-1/3}h^{i},\text{ \ \ \ \ \ \ \ \ \ \ }
a_{ij}\rightarrow 4.6^{-1/3}G_{ij}.
\nonumber
\end{equation}
}

\bea
\hat{\mathbf{e}}^{-1}\hat{\mathcal{L}_{5}}=\frac{1}{2}\hat{R}-\frac{1}{2}
G_{ij}\partial _{\hat{m}}h^{i}\partial ^{\hat{m}}h^{j}-\frac{1}{4}G_{ij}
({\mathcal{F}}^i)_{\hat{m}\hat{n}}({\mathcal{F}}^{j})^{\hat{m}\hat{n}}
\nonumber \\
+\frac{\hat{\mathbf{e}}^{-1}}{48}\,C_{ijk}\epsilon ^{\hat{n}_1\hat{n}_2
\hat{n}_3\hat{n}_4\hat{n}_5}({\mathcal{F}}^i)_{\hat{n}_1 \hat{n}_2}
({\mathcal{F}}^j)_{\hat{n}_3\hat{n}_4}(\mathcal{A}^k)_{\hat{n}_5} \ .
\label{fivelag}
\eea

Here $\hat{\mathbf{e}}$ is the determinant of the f\"{u}nfbein and $\hat{R}$
the space-time Ricci scalar, $C_{ijk}$ are real constants, symmetric in $i,j,k$. 
All the physical quantities of the theory are
determined in terms of a homogeneous cubic polynomial $\mathcal{V}$ which
defines very special geometry,\ 
\begin{eqnarray}
G_{ij} &=&-{\frac{1}{2}}{\frac{\partial }{\partial h^{i}}}{\frac{\partial }
{\partial h^{j}}}(\ln \mathcal{V})|_{\mathcal{V}=1}={\frac{9}{2}}h_{i}h_{j}-
{\frac{1}{2}}C_{ijk}h^{k}\qquad  \nonumber \\
\end{eqnarray}
where 
\begin{equation}
\mathcal{V}={\frac{1}{6}}C_{ijk}h^{i}h^{j}h^{k}=h^{i}h_{i}=1 \ ,\ \  \qquad h_i \equiv {1 \over 6} C_{ijk} h^j h^k \ .
\end{equation}
In particular we have the relation
\begin{equation}
G_{ij}h^{j}={\frac{3}{2}}h_{i} \ .
\end{equation}
The reduction ansatz is given by \cite{mohaupt3}:
\begin{equation}
\label{ansatz}
{\hat{e}}^{a}=e^{-\phi /2}e^{a},\qquad {\hat{e}}^{0}=e^{\phi }(dt-A^{0}) \ .
\end{equation}
Here $\hat{e}$ are the f\"unfbeins, $e^a$
are the vielbeins, $A^{0}$ and $\phi $ are, respectively, a gauge
field and a scalar field. All fields are independent of the 
coordinate $t$, and $e^a_t=0, A^0_t=0$.
The five dimensional flat metric is denoted by 
$\eta _{\hat{m}\hat{n}}=\left( -\epsilon ,+,+,+,\epsilon \right) ~$ while the four
dimensional one is denoted by $\eta _{ab}=\left( +,+,+,\epsilon \right)$; 
Roman indices $m,n$ denote $D=5$ frame indices, whereas $a,b...$ are 
$D=4$ frame indices.
Here $\epsilon =-1$ for reduction on a space-like direction and $\epsilon =1$
for reduction on a time-like direction. 

Note that the non-vanishing components of the $D=5$ spin connection ${\hat{\omega}}$, written in the frame basis,
are given by
\bea
{\hat{\omega}}_{0,0{\hat{a}}} &=& - \epsilon e^{\phi \over 2} \partial_a \phi
\nonumber \\
{\hat{\omega}}_{0,{\hat{a}}{\hat{b}}} &=& -{\epsilon \over 2} e^{2 \phi} (F^0)_{ab}
\nonumber \\
{\hat{\omega}}_{{\hat{a}},0{\hat{b}}} &=& -{\epsilon \over 2} e^{2 \phi} (F^0)_{ab}
\nonumber \\
{\hat{\omega}}_{{\hat{a}},{\hat{b}}{\hat{c}}} &=& e^{\phi \over 2} \bigg( \omega_{a, bc} +{1 \over 2} \eta_{ac} \partial_b \phi -{1 \over 2} \eta_{ab}
\partial_c \phi \bigg)
\eea
where indices on the LHS are $D=5$ frame indices, taken with respect to the basis ${\hat{e}}$, whereas the indices
on the RHS are $e^a$ frame indices, and $F^0 = dA^0$. The spin connection associated with the $D=4$ basis
$e^a$ has components $\omega_{a,bc}$.

The $D=5$ gauge potentials ${\cal{A}}^i$ (${\cal{F}}^i = d {\cal{A}}^i$) are decomposed as
\bea
{\cal{A}}^i = x^i (dt - A^0) + A^i, \qquad A^i_t=0
\eea
where $A^i$ are the $D=4$ gauge potentials; the scalar fields $x^i$ and gauge potentials $A^i$
are also independent of $t$.
So the components of the $D=5$ gauge field strengths ${\cal{F}}^i$ in the frame basis are given by
\bea
{\cal{F}}^i_{0 {\hat{a}}} &=& -e^{-{\phi \over 2}} \partial_a x^i
\nonumber \\
{\cal{F}}^i_{{\hat{a}} {\hat{b}}} &=& e^\phi (F^i - x^i F^0)_{ab}
\eea
where $F^i=dA^i$, and on the LHS, the indices are frame indices defined with respect to
({\ref{ansatz}}), and on the RHS $e^a$ frame indices are used.

Then, after performing the redefinitions : 
\begin{equation}
\qquad h^{i}\ =e^{-\phi }y^{i}~,\quad \quad G_{ij}\ =-2\epsilon
g_{ij}e^{2\phi }\,,
\end{equation}
and rescaling the $D=4$ gauge fields $F^0$ and $F^i$ by a factor of $\sqrt{2},$ we obtain from 
(\ref{fivelag})

\begin{eqnarray}
\mathbf{e}^{-1}\mathcal{L} &=&\frac{1}{2}R-g_{ij}\left( \partial _{a
}x^{i}\partial ^{a }x^{j}-\epsilon \partial _{a}y^{i}\partial ^{a}y^{j}\right)  \notag \\
&&+Cyyy\left[ \frac{\epsilon }{24}F^{0} \cdot F^{0}+\epsilon \frac{1}{6}\left(
gxxF^{0} \cdot F^{0}+\,g_{ij}\,F^{i} \cdot F^{j}-2\,\left( gx\right)
_{i}\!F^{i} \cdot F^{0}\right) \right]  \notag \\
&&+\frac{1}{12}\left[ 3\left( Cx\right) _{ij}F^{i} \cdot \tilde{F}^{j}-3\left(
Cxx\right) _{i}F^{i} \cdot \tilde{F}^{0}+\left( Cxxx\right) F^{0} \cdot \tilde{F}^{0}
\right]  \label{reducedaction}
\end{eqnarray}
where $R$ is the Ricci scalar of the $D=4$ manifold with metric $ds^2_4 = \delta_{ab} e^a e^b$.
We have used the notation

\begin{equation}
Chhh=C_{ijk}h^{i}h^{j}h^{k},\text{ \ \ }\left( Chh\right)
_{i}=C_{ijk}h^{i}h^{j},\text{ \ \ }\left( Cy\right) _{ij}=C_{ijk}h^{i}
\end{equation}
and $F \cdot F=F_{ab} F^{ab}$. The dual field strength is 
$\tilde{F}_{ab }=\frac{\epsilon }{2} \epsilon _{abcd}F^{cd}$, and we remark that the relationship between the $D=5$ and $D=4$ volume forms
is {\footnote{This is  the opposite sign convention to that used in \cite{mohaupt3}.}}
\bea
{\widehat{\rm dvol_5}} = - e^{-2 \phi } {\hat{e}}^0 \wedge {\rm dvol_4}
\eea
where ${\rm dvol_4}$ is the volume form of the D=4 manifold with metric $ds_4^2$.

 The explicit form of $g_{ij}$ is 
\begin{equation}
g_{ij}=\epsilon \,\frac{3}{2}\left( \frac{(Cy)_{ij}}{Cyyy}-\frac{3}{2}
\frac{(Cyy)_{i}(Cyy)_{j}}{(Cyyy)^{2}}\right) .
\end{equation}

For both values of $\epsilon$, it was demonstrated in \cite{mohaupt3} that 
(\ref{reducedaction}) can be described by the Lagrangian of the four-dimensional 
$N=2$ supergravity theory coupled to vector multiplets \cite{Andrian, craps,
vanparis}

\begin{equation}
\mathbf{e}^{-1}\mathcal{L}=\frac{1}{2}R-g_{ij}\partial _{\mu }z^{i}\partial
^{\mu }\bar{z}^{j}+\frac{1}{4}{\rm{Im}}\mathcal{N}_{IJ}F^{I}\cdot F^{J}+
\frac{1}{4}{\rm{Re}}\mathcal{N}_{IJ}F^{I}\cdot \tilde{F}^{J}.
\label{fouraction}
\end{equation}
with the cubic prepotential
\begin{equation}
F=\frac{1}{6}C_{ijk}\frac{X^{i}X^{j}X^{k}}{X^{0}} \ .
 \label{pre}
\end{equation}

It should be mentioned that the dimensional reduction of (\ref{fivelag}) on
a space-like circle was considered before in \cite{GST}. The coupling of $N=2$
vector multiplets to $N=2$ supergravity is encoded in a holomorphic
homogenous prepotential $F(X)$ of degree two.  To demonstrate the
equivalence of the reduced theory with the one given by ({\ref{fouraction}}), ({\ref{pre}}), 
the so-called $\epsilon$-complex coordinates 
($X^{I}={\rm{Re}}X^{I}+i_{\epsilon }{\rm{Im}}X^{I})$ were introduced and $F$  is taken
to be $\epsilon $-holomorphic, i.e. it depends on $\epsilon $-complex scalar fields.
Here $i_{\epsilon }$ satisfies $i_{\epsilon }=e,$ for $\epsilon =1$ and 
$i_{\epsilon }=i,$ for $\epsilon =-1$. In the symplectic formulation of the
theory, one introduces the symplectic vectors

\begin{equation}
V=\left( 
\begin{array}{c}
X^{I} \\ 
F_{I}
\end{array}
\right)
\end{equation}
satisfying the symplectic constraint

\begin{equation}
i_{\epsilon }\left( \bar{X}^{I}F_{I}-X^{I}\bar{F}_{I}\right) =-N_{IJ}X^{I}
\bar{X}^{J}=1  \label{symp}
\end{equation}
where

\begin{equation}
N_{IJ}=-i_{\epsilon }\left( F_{IJ}-\bar{F}_{IJ}\right) .
\end{equation}
$F_{I}=\frac{\partial F}{\partial X^{I}}$ and $F_{IJ}=\frac{\partial ^{2}F}
{\partial X^{I}\partial X^{J}}.$ The constraint (\ref{symp}) can be solved by
setting

\begin{equation}
X^{I}=e^{K(z,\bar{z})/2}X^{I}(z)
\end{equation}
where $K(z,\bar{z})$ is the K\"{a}hler potential. Then we have

\begin{equation}
e^{-K(z,\bar{z})}=-N_{IJ}X^{I}(z)\bar{X}^{J}(\bar{z}) \ .
\end{equation}
The resulting geometry of the the physical scalar fields $z^{i}$ of the
vector multiplets is then given by a special K\"ahler manifold with K\"{a}hler
metric 
\begin{equation}
g_{i {\bar{j}}}=\frac{\partial ^{2}K(z,\bar{z})}{\partial z^{i}\,\partial {\bar{z}}^{j}}.
\end{equation}
A convenient choice of inhomogeneous coordinates $z^{i}$ are the 
\textit{special} coordinates defined by

\begin{equation*}
X^{0}(z)=1,\text{ \ \ \ }X^{i}(z)=z^{i} \ .
\end{equation*}
The gauge field coupling matrix

\begin{equation}
\mathcal{\bar{N}}_{IJ}=F_{IJ}(X)+i_{\epsilon }\epsilon \frac{(N\bar{X})_{I}
(N\bar{X})_{J}}{\bar{X}N\bar{X}}\;.  \label{gaugemetric}
\end{equation}
For theories with cubic prepotentials in ({\ref{pre}}), we obtain
\begin{equation}
g_{ij}=\epsilon \left( \frac{3}{2}\frac{\left( Cy\right) _{ij}}{Cyyy}-\frac{9}{4}
\frac{\left( Cyy\right) _{i}\left( Cyy\right) _{j}}{(Cyyy)^{2}}\right)
\end{equation}
and

\begin{eqnarray}
\mathcal{N}_{00} &=&\frac{1}{3}Cxxx+\epsilon i_{\epsilon }Cyyy\,\left( \frac{2}{3}gxx+\frac{1}{6}\right) ,  \notag \\
\mathcal{N}_{0i} &=&-\frac{1}{2}\left( Cxx\right) _{i}-\frac{2}{3}\epsilon
i_{\epsilon }\,Cyyy\left( gx\right) _{i},\;  \notag \\
\mathcal{N}_{ij} &=&\left( Cx\right) _{ij}+\frac{2}{3}\epsilon i_{\epsilon
}g_{ij}\,Cyyy\,.\;  \label{ncomp}
\end{eqnarray}
Therefore the kinetic term of the scalar fields agrees with the reduced
theory where
\begin{equation}
z^{i}=x^{i}-i_{\epsilon }y^{i} \ .
\end{equation}
Using (\ref{ncomp}) then the gauge part of the action (\ref{fouraction})
gives

\begin{eqnarray}
&&\frac{1}{6}\epsilon Cyyy\left( \frac{1}{4}F^{0} \cdot F^{0}+gxxF^{0} \cdot F^{0}-2\left(
gx\right) _{i}F^{i} \cdot F^{0}+g_{ij}F^{i} \cdot F^{j}\right)  \notag \\
&&+\frac{1}{12}\left( CxxxF^{0} \cdot \tilde{F}^{0}-3\left( Cxx\right) _{i}F^{i}
\cdot \tilde{F}^{0}+3\left( Cx\right) _{ij}F^{i} \cdot \tilde{F}^{j}\right)
\end{eqnarray}
which agrees with the the reduced Lagrangian.

\section{Reduced Killing Spinor Equations}

In this section we start with the supersymmetry variation of the gravitini
and gaugino in the five dimensional supergravity theory and reduce them to
four dimensions. The associated Killing spinor equations are
\bea
\bigg( {{{\hat{D}}}}_{\hat{m}} + {i \over 8} h_i \big( \Gamma_{\hat{m}}{}^{{\hat{n}}_1 {\hat{n}}_2}
-4 \delta_{{\hat{m}}}^{{\hat{n}}_1} \Gamma^{{\hat{n}}_2} \big) {\cal{F}}^i_{{\hat{n}}_1 {\hat{n}}_2} \bigg) {\hat{\varepsilon}} =0
\label{grav}
\eea
and
\bea
\bigg( \big({\cal{F}}^i - h^i h_j {\cal{F}}^j \big)_{{\hat{n}}_1 {\hat{n}}_2} \Gamma^{{\hat{n}}_1 {\hat{n}}_2}
+2i {\hat{\nabla}}_{\hat{m}} h^i \Gamma^{\hat{m}} \bigg)  {\hat{\varepsilon}} =0 \ .
\label{dil}
\eea
Here ${{\hat{D}}}_{\hat{m}}=\partial _{\hat{m}}+\frac{1}{4}
{\hat{\omega}}_{{\hat{m}}, {\hat{n}}_1 {\hat{n}}_2} \Gamma^{{\hat{n}}_1 {\hat{n}}_2}$ is the five dimensional covariant derivative.
Note that $\Gamma_0$ squares to $-\epsilon$, and $\Gamma^0 = -\epsilon \Gamma_0$.
We first reduce ({\ref{grav}}) and ({\ref{dil}}) to $D=4$; throughout what follows the rescaling of the $D=4$ gauge field strengths
by $\sqrt{2}$ is taken into account.

First consider the ${\hat{m}}=0$ component of ({\ref{grav}}); this reduces from $D=5$ to $D=4$ to give
\bea
\label{reduc1}
\bigg( {i \over 2} e^{\phi \over 2} h_i \Gamma^a \big( \partial_a x^i + i \partial_a y^i \Gamma_0 \big)
+{i \over 4 \sqrt{2}} e^{2 \phi} \Gamma^{ab} \big( h_i \Gamma_0 (F^i-x^i F^0)_{ab}+i \epsilon e^\phi F^0_{ab} \big) \bigg)
{\hat{\varepsilon}} =0 \ .
\eea
Consider also the reduction of the $D=5$ gaugino equation ({\ref{dil}}); which gives
\bea
\label{reduc2}
\bigg( -{1 \over \sqrt{2}} e^{3 \phi \over 2} \Gamma_0 (\delta^i_j - h^i h_j) (F^j-x^j F^0)_{ab} \Gamma^{ab} \qquad \qquad
\nonumber \\
+ \Gamma^a  \big( \partial_a x^i - h^i h_j \partial_a x^j +i \Gamma_0 \partial_a y^i
-i e^\phi h^i \partial_a \phi \Gamma_0 \big) \bigg) {\hat{\varepsilon}} =0 \ .
\eea

After some calculation, details of which are given in Appendix A, the two conditions ({\ref{reduc1}})
and ({\ref{reduc2}}) can be combined into the following expression:
\bea
\label{combin1}
\frac{i}{2}e^{K/2}(\text{Im}\,\mathcal{N})_{IJ}\Gamma^{ab} F^J_{ab}
\left[ \text{Im}(g^{i\bar{j}}\mathcal{D}_{\bar{j}}{\bar{X}}^{I})+
i\epsilon \Gamma_0 \text{Re}(g^{i\bar{j}}\mathcal{D}_{\bar{j}}{\bar{X}}^{I})\right] 
{\hat{\varepsilon}} 
\nonumber \\
+\Gamma^a \partial_a \left[ \text{Re}z^{i}-i\Gamma _0
\text{Im}z^{i}\right] {\hat{\varepsilon}}  =0
\eea
where 
\begin{equation}
\mathcal{D}_{^{\bar{j}}}\bar{X}^{I}=\mathcal{\partial }_{^{\bar{j}}}\bar{X}^{I}+\mathcal{\partial }_{^{\bar{j}}}K\bar{X}^{I} \ .
\end{equation}

In particular, one finds that ({\ref{reduc1}}) is obtained from ({\ref{combin1}}) by contracting with $h_i$,
whereas one obtains ({\ref{reduc2}}) by considering the directions of ({\ref{combin1}}) which are orthogonal to $h_i$.

Next consider the ${\hat{m}}={\hat{a}}$ component of ({\ref{grav}});
this reduces to $D=4$ to give the following expression:
\bea
\label{reduc3a}
{D}_a {\hat{\varepsilon}} +
\bigg( {1 \over 2 \sqrt{2}} e^{3 \phi \over 2} \Gamma_0 \Gamma^b (F^0)_{ab}-{1 \over 4} \Gamma_a{}^b \partial_b \phi
-{i \over 4} \epsilon \Gamma_0 \Gamma_a{}^b e^{-\phi} h_i \partial_b x^i +{i \over 2} \epsilon \Gamma_0 e^{-\phi} h_i \partial_a x^i 
\nonumber \\
+{i \over 4 \sqrt{2}} h_i \Gamma_a{}^{bc} e^{\phi \over 2} (F^i-x^i F^0)_{bc}
-{i \over \sqrt{2}} h_i e^{\phi \over 2} (F^i-x^i F^0)_{ab} \Gamma^b \bigg) {\hat{\varepsilon}} =0 \ . 
\nonumber \\
\eea
In order to rewrite this expression, we introduce the $U(1)$ (para)-K\"ahler potential{\footnote{To be distinguished from the 
gauge potentials $A^0, A^i$}}
\bea
\label{pkpot}
A_a = -{i_\epsilon \over 2} \big( \partial_i K \partial_a z^i
- \partial_{{\bar{i}}} K \partial_a z^{\bar{i}} \big) \ .
\eea
This can be recast as
\bea
A_a = -{3 \over 2} e^{-\phi} h_i \partial_a x^i \ .
\eea

Then, using the identities listed in Appendix A, ({\ref{reduc3a}}) can be rewritten as
\bea
\label{reduc3b}
{D}_a {\hat{\varepsilon}}
+ \bigg( {1 \over 4} \partial_a \phi -{\frac{i}{2}}A_a \epsilon \Gamma_0  +\frac{i}{4}e^{{K \over 2}}
\Gamma^{bc} F^{I}_{bc} \big( {\rm{Im}}X^{J}+i\epsilon \Gamma_0 {\rm{Re}}
X^{J}\big) (\text{Im}\,\mathcal{N})_{IJ}\Gamma_a  \bigg) {\hat{\varepsilon}}
\nonumber \\
+{1 \over 2} \epsilon \Gamma_a \Gamma_0 e^{-{3 \phi \over 2}}
\bigg( {i \over 4 \sqrt{2}} e^{2 \phi} \Gamma^{bc} (h_i \Gamma_0 (F^i-x^i F^0)_{bc}
+i \epsilon e^{\phi} F^0_{bc})
\nonumber \\
+{i \over 2} e^{\phi \over 2} h_i \Gamma^b (\partial_b x^i +i \partial_b y^i \Gamma_0) \bigg)
{\hat{\varepsilon}} =0 \ .
\nonumber \\
\eea
Observe that the second and third line of this expression can be removed by
using  ({\ref{reduc1}}) and so on setting
\bea
{\hat{\varepsilon}} = e^{-{\phi \over 4}} \varepsilon
\eea
it follows that ({\ref{grav}}) and ({\ref{dil}}) can be rewritten as
\bea
\label{eq1}
{D}_a \varepsilon -{\frac{i}{2}}  \epsilon A_a  \Gamma_0 \varepsilon
+  \frac{i}{4}e^{{K \over 2}}
\Gamma^{bc} F^{I}_{bc} \big( {\rm{Im}}X^{J}+i\epsilon \Gamma_0 {\rm{Re}}
X^{J}\big) (\text{Im}\,\mathcal{N})_{IJ}\Gamma_a   \varepsilon =0
\eea
and
\bea
\label{eq2}
\frac{i}{2}e^{K/2}(\text{Im}\,\mathcal{N})_{IJ}\Gamma^{ab} F^J_{ab}
\left[ \text{Im}(g^{i\bar{j}}\mathcal{D}_{\bar{j}}{\bar{X}}^{I})+
i\epsilon \Gamma_0 \text{Re}(g^{i\bar{j}}\mathcal{D}_{\bar{j}}{\bar{X}}^{I})\right] 
\varepsilon
\nonumber \\
+\Gamma^a \partial_a \left[ \text{Re}z^{i}-i\Gamma _{0}
\text{Im}z^{i}\right] \varepsilon  =0 \ .
\eea

\section{Chiral Decomposition}

In this section we express the transformations (\ref{eq1}) and ({\ref{eq2}}) in terms of
chiral spinors. In order to define the various projections, it is convenient to note that
{\footnote{We remark that the sign in ({\ref{vcon}}) is fixed by requiring that the integrability 
conditions of the Killing spinor equations ({\ref{grav}}) and ({\ref{dil}}) should be
consistent with the gauge field equations obtained from ({\ref{fivelag}}).}}
\bea
\label{vcon}
\Gamma_{{\hat{n}}_1 {\hat{n}}_2 {\hat{n}}_3
{\hat{n}}_4 {\hat{n}}_5} = i ({\widehat{{\rm dvol}_5}})_{{\hat{n}}_1 {\hat{n}}_2 {\hat{n}}_3
{\hat{n}}_4 {\hat{n}}_5}
\eea
which implies that
\bea
\label{gprj}
\Gamma_0 \Gamma_{ab} = {i \over 2} \epsilon_{ab}{}^{cd} \Gamma_{cd}\ .
\eea

In the Minkowski case ($\epsilon=-1$), we decompose the spinor $\varepsilon $
in terms of chiral spinors as $\varepsilon =\varepsilon _{-}+\varepsilon
_{+},$ where we set
\begin{eqnarray}
\Gamma _{\pm } &=&\frac{1}{2}\left( 1\pm \Gamma_0 \right)
\nonumber  \\
\Gamma_{\pm} \varepsilon_\pm &=& \varepsilon_\pm
\nonumber \\
\Gamma_{\pm} \varepsilon_\mp &=& 0
\end{eqnarray}
and we also define
\begin{equation}
F_{ab }^{\pm }=\frac{1}{2}\left( F_{ab}\pm i\tilde{F}_{ab}\right) \ .
\end{equation}
Also, ({\ref{gprj}}) implies that 
\begin{equation}
\Gamma .F=\Gamma .\left( F^{-}\ \Gamma _{+}+F^{+}\ \Gamma _{-}\right) \ .
\label{nicerelation}
\end{equation}
We find that ({\ref{eq1}}) and ({\ref{eq2}}) can be rewritten as
\bea
{D}_a \varepsilon_\pm \pm {i \over 2} A_a \varepsilon_\pm
\pm {1 \over 4} e^{K \over 2} \Gamma^{bc} F^{\mp I}_{bc} \big( {\rm Re} X^J \pm i {\rm Im} X^J \big)
{\rm Im} \ {\mathcal{N}}_{IJ} \Gamma_a \varepsilon_\mp =0
\eea
and
\bea
\pm {1 \over 2} e^{K \over 2} {\rm Im} \ {\mathcal{N}}_{IJ} \Gamma^{ab} F^{\mp J}_{ab}
\big( {\rm Re} (g^{i {\bar{j}}} {\mathcal{D}}_{\bar{j}} {\bar{X}}^I) \pm  i 
 {\rm Im} (g^{i {\bar{j}}} {\mathcal{D}}_{\bar{j}} {\bar{X}}^I) \big) \varepsilon_\pm
 \nonumber \\
 + \Gamma^a \partial_a \big( {\rm Re} z^i \pm i {\rm Im} z^i \big) \varepsilon_\mp =0 \ .
 \eea

This is in agreement with the Killing spinor equations given by \cite{vanparis}
(on making the identification $\varepsilon^1= \varepsilon_+, \ \varepsilon_2 = \varepsilon_-$):

\begin{eqnarray}
D_{a }\varepsilon ^{\alpha }+
\frac{i}{2}A_a \varepsilon ^{\alpha }+\frac{1}{4}(\text{Im}\,\mathcal{N})_{IJ}
X^{J}(z)e^{K/2}\Gamma .F^{-I}\epsilon ^{\alpha \beta }\Gamma_a \epsilon _{\beta } = 0
 \notag \\
-\frac{1}{2}e^{{K \over 2}}(\text{Im}\,
\mathcal{N})_{IJ}g^{i\bar{j}}\mathcal{D}_{^{\bar{j}}}\bar{X}^{I}(\bar{z}
)\gamma .F^{-J}\epsilon _{\alpha \beta }\epsilon^{\beta }+\Gamma^a
\partial_a z^{i}\epsilon _{\alpha } = 0 \ .
 \label{chiral}
\end{eqnarray}

Next consider the Euclidean case ($\epsilon = 1$). There are two alternative chiral decomposition possible.
For the first, we define
\begin{eqnarray}
\Gamma _{\pm } &=&\frac{1}{2}\left( 1\pm i \Gamma_0 \right)
\nonumber  \\
\Gamma_{\pm} \varepsilon_\pm &=& \varepsilon_\pm
\nonumber \\
\Gamma_{\pm} \varepsilon_\mp &=& 0
\end{eqnarray}
with
\begin{equation}
F_{ab}^{\pm}=\frac{1}{2}\left( F_{ab} \pm \tilde{F}_{ab}\right) \ ,
\end{equation}
and ({\ref{gprj}}) implies that
\begin{equation}
\Gamma .F=\Gamma .\left( F^{-}\ \Gamma _{+}+F^{+}\ \Gamma _{-}\right)  \ .
\label{nicerelation2}
\end{equation}
We find that ({\ref{eq1}}) and ({\ref{eq2}}) can be rewritten as
\bea
{D}_a \varepsilon_\pm \mp {1 \over 2} A_a \varepsilon_\pm
\pm {i \over 4} e^{K \over 2} \Gamma^{bc} F^{\mp I}_{bc} \big( {\rm Re} X^J \pm  {\rm Im} X^J \big)
{\rm Im} \ {\mathcal{N}}_{IJ} \Gamma_a \varepsilon_\mp =0
\eea
and
\bea
\pm {i \over 2} e^{K \over 2} {\rm Im} \ {\mathcal{N}}_{IJ} \Gamma^{ab} F^{\mp J}_{ab}
\big( {\rm Re} (g^{i {\bar{j}}} {\mathcal{D}}_{\bar{j}} {\bar{X}}^I) \pm   
 {\rm Im} (g^{i {\bar{j}}} {\mathcal{D}}_{\bar{j}} {\bar{X}}^I) \big) \varepsilon_\pm
 \nonumber \\
 + \Gamma^a \partial_a \big( {\rm Re} z^i \pm  {\rm Im} z^i \big) \varepsilon_\mp =0 \ .
 \eea
This is the form of the Killing spinor equations expressed in terms of
the so-called {\it adapted coordinates} \cite{mohaupt3}.

For the second chiral decomposition in the Euclidean case, we define \cite{mohaupt1}
\begin{eqnarray}
\Gamma _{\pm } &=&\frac{1}{2}\left( 1\pm i e \Gamma_0 \right)
\nonumber  \\
\Gamma_{\pm} \varepsilon_\pm &=& \varepsilon_\pm
\nonumber \\
\Gamma_{\pm} \varepsilon_\mp &=& 0
\end{eqnarray}
and let
\begin{equation}
F_{ab}^{\pm}=\frac{1}{2}\left( F_{ab} \pm e \tilde{F}_{ab}\right) \ .
\end{equation}
With these conventions, ({\ref{nicerelation2}}) holds,
and we find that ({\ref{eq1}}) and ({\ref{eq2}}) can be rewritten as
\bea
{D}_a \varepsilon_\pm \mp {e \over 2} A_a \varepsilon_\pm
\pm {ie \over 4}  e^{K \over 2} \Gamma^{bc} F^{\mp I}_{bc} \big(  {\rm Re} X^J \pm e {\rm Im} X^J \big)
{\rm Im} \ {\mathcal{N}}_{IJ} \Gamma_a \varepsilon_\mp =0
\eea
and
\bea
\pm {i e\over 2}  e^{K \over 2} {\rm Im} \ {\mathcal{N}}_{IJ} \Gamma^{ab} F^{\mp J}_{ab}
\big( {\rm Re} (g^{i {\bar{j}}} {\mathcal{D}}_{\bar{j}} {\bar{X}}^I) \pm   
e {\rm Im} (g^{i {\bar{j}}} {\mathcal{D}}_{\bar{j}} {\bar{X}}^I) \big) \varepsilon_\pm
 \nonumber \\
 + \Gamma^a \partial_a \big( {\rm Re} z^i \pm  e {\rm Im} z^i \big) \varepsilon_\mp =0 \ .
 \eea

\section{Discussion}

In this letter we have derived the Killing spinor equations for Euclidean supergravity 
theories coupled to Abelian vector multiplets ({\ref{eq1}}) and ({\ref{eq2}}).
We have obtained the four-dimensional Killing spinor equations from the
reduction of those in the five-dimensional theory. We explicitly show
how this is achieved by writing the reduced equations in an $\epsilon$-K\"ahler covariant
formalism. These equations were also rewritten, for the Euclidean case, in terms of
chiral spinors using both the adapted and para-complex co-ordinates.
$\epsilon$-special K\"ahler geometry in Euclidean theories is expected to play an 
important role in the analysis of instantons, solitons and cosmological solutions
in supergravity and M-theory.
The Killing spinor equations given in 
({\ref{eq1}}) and ({\ref{eq2}}) provide the starting point to find general instanton solutions of
the effective Euclidean $N=2$ supergravity action coupled to $N=2$ matter
multiplets. As for the case of black holes, one also expects that the rich geometric structure of
the theory will lead to a simplified approach for finding new instanton solutions.

Spinorial geometry techniques \cite{papadopoulos} have proven to be
a very useful tool in finding all instanton solutions preserving various
fractions of supersymmetry. Those techniques were also used recently in
finding solutions of Einstein-Maxwell theory with \cite{instantons1} or without \cite{instantons2}
a cosmological constant, as well as the supersymmetric solutions of euclidean $N=4$ super
Yang-Mills theory \cite{instantonsklemm};
where interesting relations to integrable models 
\cite{instantons1} and the Hitchin equations \cite{instantonsklemm} were found. We will
report on the instanton solutions with vector multiplets in a separate publication. Another direction which needs
to be investigated is the construction of gauged Euclidean supergravity models.

\bigskip

\setcounter{section}{0}\setcounter{equation}{0}

\appendix{$\epsilon$-K\"ahler Special Geometry Identities}

In this Appendix, we summarize a number of useful identities.
First, consider rewriting the reduction of ({\ref{reduc1}}) and ({\ref{reduc2}})
in a special $\epsilon$-K\"ahler covariant fashion as ({\ref{combin1}}).

This is done by making use of the following identities:
\bea
\label{id1}
{\rm Im} \ {\cal{N}}_{I0} 
\left[ \text{Im}(g^{i\bar{j}}\mathcal{D}_{\bar{j}}{\bar{X}}^{I})+
i\epsilon \Gamma_0 \text{Re}(g^{i\bar{j}}\mathcal{D}_{\bar{j}}{\bar{X}}^{I})\right] 
&=& \big(2 \epsilon e^{4 \phi} + 2i e^{3 \phi} \Gamma_0 h_j x^j \big) h^i
\nonumber \\
&-&4i e^{3 \phi} \Gamma_0 (x^i - h_j x^j h^i)
\eea
and
\bea
\label{id2}
{\rm Im} \ {\cal{N}}_{I\ell}  \left[ \text{Im}(g^{i\bar{j}}\mathcal{D}_{\bar{j}}{\bar{X}}^{I})+
i\epsilon \Gamma_0 \text{Re}(g^{i\bar{j}}\mathcal{D}_{\bar{j}}{\bar{X}}^{I})\right] 
= -2i e^{3 \phi} \Gamma_0 h^i h_\ell +4i e^{3 \phi} \Gamma_0 (\delta^i_\ell - h^i h_\ell)
\nonumber \\
\eea
where we have also used the identities
\bea
{\mathcal{D}}_{\bar{j}} {\bar{X}}^0 &=& -{3 \over 2} \epsilon i_\epsilon e^{-\phi} h_j
\nonumber \\
{\mathcal{D}}_{\bar{j}} {\bar{X}}^i &=& \delta^i_j -{3 \over 2} h^j h_i -{3 \over 2} \epsilon i_\epsilon e^{-\phi} h_j x^i
\eea
and
\bea
g^{ij} h_j =-{2 \over 9} \epsilon e^{-\phi} h^i Cyyy 
\eea
and
\bea
Cyyy = 6 e^{3 \phi} , \qquad e^{-K} = {4 \over 3} Cyyy = 8 e^{3 \phi} \ .
\eea
Another useful identity used to obtain ({\ref{combin1}}) is
\bea
\Gamma^a \partial_a \big( {\rm Re} \ z^i - i \Gamma_0 {\rm Im} \ z^i \big)
&=& \Gamma^a \bigg( \partial_a (x^i+i \Gamma_0 y^i) -h^i h_j \partial_a x^j -i \Gamma_0 h^i \partial_a \phi e^\phi \bigg)
\nonumber \\
&+& \Gamma^a h^i (h_j \partial_a x^j + i \Gamma_0 \partial_a \phi e^\phi)
\eea
where the expression on the first line of the RHS is projected orthogonal to the direction of $h_i$,
and the second line contains the term parallel to $h_i$.

A number of useful identities used to obtain ({\ref{reduc3b}}) are 
\bea
{i \over 4} e^{K \over 2} \left[ {\rm Im} \ X^J + i \epsilon \Gamma_0 {\rm Re} \ X^J \right] {\rm Im} \ {\mathcal{N}}_{IJ}
&=& {i \over 8 \sqrt{2}} e^{-{3 \phi \over 2}} \big(-i \epsilon \Gamma_0
{\rm Im} \ {\mathcal{N}}_{I0} 
\nonumber \\
&-& (i \epsilon \Gamma_0 x^j + y^j) {\rm Im} \ {\mathcal{N}}_{Ij} \big)
\nonumber \\
\eea
and
\bea
-i \epsilon \Gamma_0 {\rm Im} \  {\mathcal{N}}_{00} - (i \epsilon \Gamma_0 x^j + y^j) {\rm Im} \ {\mathcal{N}}_{0j}
= 6 e^{3 \phi} \big(-{i \over 6} \Gamma_0 -{1 \over 2} e^{-\phi} h_i x^i \big)
\eea
and
\bea
-i \epsilon \Gamma_0 {\rm Im} \ {\mathcal{N}}_{i0} - (i \epsilon \Gamma_0 x^j + y^j) {\rm Im} \ {\mathcal{N}}_{ij}
= 3 e^{2 \phi} h_i
\eea
together with
\bea
(gy)_i = -{3 \over 4} \epsilon e^{-\phi} h_i \ .
\eea

\vskip 0.5cm
\noindent{\bf Acknowledgements} \vskip 0.1cm
\noindent  WS and JG would like to thank the Isaac Newton Institute Cambridge
for support during this work. JG is supported by the STFC grant, ST/1004874/1.

\end{document}